\documentstyle[12pt]{article}

\begin{document}
\thispagestyle{empty}

\mbox{}
\vspace{1cm}
\begin{center}
{\bf DISTRIBUTIONS OF ABSOLUTE CENTRAL MOMENTS} \\
\vspace{0.5cm}
{\bf FOR RANDOM WALK SURFACES} \\
\vspace{0.5cm}
{\it A. J. McKane}$^{\dag}$ {\it and R. K. P. Zia}$^{\ddag}$ \\
\vspace{0.5cm}
$^{\dag}$Department of Theoretical Physics \\
University of Manchester \\ Manchester M13 9PL, UK \\

\vspace{0.5cm}

$^{\ddag}$Center for Stochastic Processes in Science and Engineering \\
and Department of Physics \\
Virginia Polytechnic and State University \\
Blacksburg, Virginia 24061, USA \\
\end{center}

\begin{abstract}
We study periodic Brownian paths, wrapped around the surface of a cylinder.
One characteristic of such a path is its width square, $w^2$, defined
as its variance. Though the average of $w^2$ over all possible
paths is well known, its full distribution function was investigated only
recently. Generalising $w^2$ to $w^{(N)}$, defined as the $N$-th
power of the {\it magnitude} of the deviations of the path from its mean,
we show that the distribution functions of these also scale and
obtain the asymptotic behaviour for both large and small $w^{(N)}$.
\end{abstract}

\newpage
\pagenumbering{arabic}

\section{Introduction}

Of all the quantities which characterise a near-planar surface,
the width, $w$, defined as the standard deviation of the position of the
surface,
is probably the most frequently studied. If this surface is a member of
a Gaussian ensemble, then $\langle w^2 \rangle$,
the ensemble average of $w^2$,
is a well-known function of the parameters of the Gaussian as well as
the
size of the surface. In particular, if the surface is a one
dimensional object of length $T$, then
$\langle w^2 \rangle \propto T$, as $T\rightarrow \infty$. In a
recent paper \cite{gof} it was pointed out that the entire {\it distribution}
of $w^2$,
rather than merely the average, would provide a better picture of the
surface. Indeed, this distribution was shown to obey scaling and the universal
scaling function was computed. An analogue in critical phenomena is the
difference between measuring a single critical exponent and finding data
collapse in a range of parameter space. It is indisputable that the latter
gives far more details of the system than the former.

In a similar vein, one may extract further information from the fluctuating
surfaces by analysing moments or cumulants other than the second. For
Gaussian
surfaces, the {\it averages} of these can be related to $\langle w^2 \rangle$,
but their {\it distributions} are less trivial. Unlike the case
for $w^2$,
the Laplace transforms of these distributions are associated with
non-Gaussian field theories. Equivalently, for one-dimensional surfaces,
we must now deal with quantum anharmonic oscillators. In this paper we
generalise the analysis begun in \cite{gof} and show that
these distributions also scale, while their universal functions can be
found in the asymptotic regimes where the argument is
small or is large.

In order to have sufficient analytical control over the fluctuations,
the one-dimensional surfaces in \cite{gof} were modelled by
Brownian paths $\{ h_t \}$
in the ``time" interval $0\le t \le T$, with periodic boundary conditions.
Thus, a surface may be thought of as a line wrapped around a cylinder
of circumference $T$. By identifying the
continuous path $h_t$ as the height of a surface, the
``width square" is a simple functional quadratic in $h_t$:
\begin{equation}
\label{2moment}
w^2[h_t] = \overline{\psi ^2_t} =
\overline{\left[ h_t - \overline{h_t} \right]^2}
\end{equation}
where
the average, $\overline f$, of a function $f$ is defined as
\begin{equation}
\label{defav}
\overline{f(h_t)}=\frac{1}{T} \int_{0}^{T} dt \,f(h_t) \, ,
\end{equation}
and where $\psi _t \equiv h_t - \overline{h_t}$ is the usual deviation from
the
``average height".
Being Brownian paths, the weight of each $h_t$ is a simple Gaussian, i.e.,
exp $\{ - \int^T_0 dt \, \left[\frac{1}{2} \dot{h}^2_t \right] \} $ ,
where $\dot{h}_t=dh_t/dt$.
For the ensemble average, denoted by $\langle ... \rangle$,
one must sum over all appropriate paths, i.e., perform functional integrals
(or path-integrals \cite{fh}) over periodic functions $h_t$.
Since none of the quantities
we encounter will depend on the overall height, we may just as well sum
over
$\{\psi_t\}$, with special attention paid to the constraint
$\int^T_0 dt \, \psi _t = 0$. With this set up, it is standard to show
that $\langle w^2 \rangle=T/12$.

The key to calculating the probability density $P(w^2;T)$ was expressing
it the form of a path-integral and noting that its Laplace transform can
be computed explicitly \cite{gof}. Beginning with
\begin{eqnarray}
\label{wiener2}
P(w^2;T) & = & \left\langle \delta \left( w^2 -
\overline{\left[ h_t - \overline{h_t} \right]^2} \ \right) \right\rangle
\nonumber \\
& = & {\cal N} \int {\cal D}[h] \, \delta \left( w^2
- \overline{\left[ h_t - \overline{h_t} \right]^2}\ \right)
\exp \left( - \,\frac{T}{2}\, \overline{\dot{h}_t^2} \, \right) \ ,
\end{eqnarray}
where ${\cal N}$ is a normalisation constant, one can easily show that
it has the scaling form
$P(w^2;T) = \langle w^2\rangle ^{-1}\Phi (x)$, where
$x$ is the scaling variable $w^2 /\langle w^2\rangle$. Inverting the
exact result for its Laplace transform
\begin{equation}
\label{g2}
G (\lambda ;T) = \frac{\sqrt{\lambda T/2}}{\sinh (\sqrt{\lambda T/2})}
\, ,
\end{equation}
the associated scaling function is obtained:
\begin{equation}
\label{phi2}
\Phi (x) = \frac{\pi ^2}{3} \sum^{\infty}_{n=1} (- 1 )^{n - 1}\, n^2 \,
\exp\left( -\frac{\pi ^2}{6} \, n^2 x \right) \, ,
\end{equation}
from which the asymptotic properties can be found.

The aim of this paper is to extend these results to arbitrary
absolute central moments:
\begin{equation}
\label{Nthmoment}
w^{(N)}[h_t] \equiv \overline{|\psi_t |^N} =
\overline{ | h_t - \overline{h_t} | ^N} \, .
\end{equation}
Notice that, for even $N$, these are precisely the
central moments associated with
a particular configuration $h_t$. Using the absolute values of $\psi$,
it is possible to study arbitrary $N>0$, and these are known as the
absolute central moments (associated with a specific path).
The outline of the paper is as follows. In section 2, a brief,
self-contained, summary of the formalism for this general case
is included. Then,
we show that $P_N (w^{(N)};T)$ does indeed scale. Analytic expressions
for
the asymptotic behavior of the scaling function are found, for large and
small arguments, in sections 3 and 4 respectively.
We conclude with some general comments in section 5.

\section{Scaling form for the probability distribution}

Following Ref. \cite {gof} , let us express the full distribution of $w^{(N)}$
as
\begin{equation}
\label{wiener}
P_N (w^{(N)};T) = {\cal N} \int {\cal D}[h] \, \delta \left( w^{(N)}
- \overline{ | h_t - \overline{h_t} | ^N}\ \right)
\exp \left( - \,\frac{T}{2}\, \overline{\dot{h}_t^2} \, \right) \ .
\end{equation}
Its Laplace transform, which generates the moments of (\ref{wiener}),
is
\begin{eqnarray}
\label{laplacet}
G_N (\lambda ;T) & = & \int_{0}^{\infty} d\zeta \,P_N(\zeta ;T)
e^{-\lambda \zeta} \nonumber \\
& = & {\cal N} \int {\cal D}[h] \,
\exp\left[- \,\frac{T}{2} \overline{\dot{h}_t^2} \, -\lambda
\overline{| h_t - \overline{h_t} |^N} \right] \, .
\end{eqnarray}
Writing this expression in terms of $\psi _t$ and using the condition
$G_N (0;T) = 1$ to determine the normalisation constant ${\cal N}$, we
obtain
\begin{equation}
\label{genfun}
G_N (\lambda ;T) = \frac{\int {\cal D}[\psi ]
\,\delta \left( \int^T_0 dt \, \psi _t \right) \exp - \int^T_0 dt \, \left[
\frac{1}{2} \dot{\psi}^2_t + \frac{\lambda}{T} |\psi_t |^N \right]}{{\int
\cal D}[\psi ]
\,\delta \left( \int^T_0 dt \, \psi _t \right) \exp - \int^T_0 dt \, \left[
\frac{1}{2} \dot{\psi}^2_t \right]}
\end{equation}
Apart from the constraint $\int^T_0 dt \, \psi _t = 0$, this is the imaginary
time propagator for a quantum-mechanical particle moving in the potential
$|\psi |^N$. As we will see in section 4, this connection will be exploited
in the study of the asymptotics of $P_N$ for small $w^{(N)}$.

To elucidate the general structure of $P_N$ and its scaling properties,
it is useful to define a dimensionless variable
$\tau = t/T$ and then to make the following rescaling:
\begin{equation}
\label{defchi}
\chi _{\tau} = T^{-1/2}\, \psi _{\tau} \, .
\end{equation}
Defining
\begin{equation}
\label{mu}
\mu = N \lambda T^{N/2} \, ,
\end{equation}
we have, for the generating function,
\begin{equation}
\label{dimform}
G_N (\lambda ;T) = \frac{\int {\cal D}[\chi ] \,\delta
\left( \int^1_0 d\tau \, \chi _{\tau} \right) \exp - \int^1_0 d\tau \,
\left[
\frac{1}{2} \dot{\chi }^2_{\tau} + \frac{\mu}{N} | \chi _{\tau} | ^N \right]}
{{\int \cal D}[\chi ] \,\delta \left(
\int^1_0 d\tau \, \chi _{\tau} \right) \exp - \int^1_0 d\tau \, \left[
\frac{1}{2} \dot{\chi }^2_{\tau} \right]}
\end{equation}
where $\dot{\chi }_{\tau}$ now means $d\chi _{\tau}/d\tau$. This expression
shows that $G_N (\lambda ;T)$ is a function of $\mu $ only, so that
we may write
\begin{equation}
\label{gG}
g_N (\mu ) \equiv G_N (\lambda ;T)
\end{equation}
Expressing $P_N (w^{(N)};T)$ as the inverse Laplace transform of $g$:
\begin{eqnarray}
\label{invlap}
P_N (w^{(N)};T) & = & \int^{i\infty}_{-i\infty} \, \frac{d\lambda}{2\pi
i}
\, g_N (N\lambda T^{N/2} ) \, \exp \, ({\lambda w^{(N)}}) \nonumber \\
& = & \frac{1}{NT^{N/2}} \int^{i\infty}_{-i\infty} \, \frac{d\mu }{2\pi
i}
\, g_N (\mu ) \, \exp \left( \frac{\mu}{N}\, \frac{w^{(N)}}{T^{N/2}} \right)
\end{eqnarray}
we see that $T^{N/2} P_N (w^{(N)};T)$ is a function of $w^{(N)}/T^{N/2}$
only. To put this in the scaling form, we simply note that
$\langle w^{(N)} \rangle$ is proportional to $T^{N/2}$:
\begin{eqnarray}
\label{meanw}
\langle w^{(N)} \rangle & = & \int^{\infty}_{0} d\zeta \, \zeta \,
P_N (\zeta ;T) \nonumber \\
& = & - \left. \frac{dG_N}{d\lambda}\right|_{\lambda = 0} \nonumber \\
& = & - NT^{N/2}\, \left. \frac{d\ }{d\mu} g_N (\mu ) \right|_{\mu = 0}
\end{eqnarray}
Using this result, (\ref{invlap}) may be written as
\begin{equation}
\label{scaling}
P_N (w^{(N)};T) = \frac {1}{\langle w^{(N)} \rangle} \, \Phi _N (x)
\end{equation}
where
\begin{equation}
\label{ex}
x \equiv w^{(N)}/\langle w^{(N)} \rangle
\end{equation}
and
\begin{equation}
\label{expphi}
\Phi_N (x) = \left| g'_N (0)\right| \, \int^{i\infty}_{-i\infty}
\, \frac{d\mu }{2\pi i}
\, g_N (\mu ) \, \exp \left( \mu x |g'_N (0)| \right)
\end{equation}

Equations (\ref{scaling})-(\ref{ex}) are an expression of the fact that
$P_N$
scales. From (\ref{expphi}) one sees that in order to find an explicit
form
for $\Phi _N (x)$, the function $g_N (\mu )$ has to be determined. For
$N \ne 2$ this can only be determined in special limits. Nevertheless,
as we shall
see in the next two sections, one can go quite a long way in finding the
limiting forms of $\Phi _N$ for large and small $x$ using path-integral
techniques.

We end this section by noting that, since we have Brownian paths, the
ensemble averages of the moments $\langle w^{(N)} \rangle$ are quite simple.
For even $N$, they are $(N-1)!! (T/12)^{N/2}$. For arbitrary $N$, we
rely on standard analytic continuation techniques to arrive at
\begin{equation}
\label{gprime}
\left| g'_N (0)\right| = \Gamma \left( \frac{N + 1}{2} \right)
/ \left\{ 6^{N/2} N \pi^{1/2} \right\} \, .
\end{equation}
However, we will continue to write $g'_N (0)$, both for convenience and
to emphasise its role.

\section{The scaling function for large x}

It is not surprising that the behaviour of $\Phi _N (x)$
for $x \gg 1$ is controlled by the singularities of $g_N (\mu )$ for
near the origin of $\mu $. In the $N=2$ case, these singularities are
simple poles on the negative real axis. In general, we expect a branch
cut
and its discontinuity near $\mu=0$ to dominate the asymptotics of $\Phi
_N$. One approach is to compute these singularities, using
techniques of asymptotic
expansion around instanton solutions \cite{Zinn}, followed by an inversion
of the Laplace transform. Alternatively, one can insert (\ref{dimform})
into
(\ref{expphi}), find the saddle point in the combined space of
$\{ \mu , \chi _{\tau} \}$ and perform the Gaussian integrations over
small variations in its neighbourhood. We shall follow the latter route,
which seems to be simpler.

As will be seen, we are able to compute explicitly only the leading
two terms in $\ln \, \Phi_N$, i.e., $x^{2/N}$ and $\ln \, x$. Thus, we
will drop all proportionality constants, for clarity, when writing $\Phi_N$.
In particular, the denominator of (\ref{dimform}) is clearly $x$
independent and will be suppressed, though its role in regulating the
Gaussian integrations is obviously essential. So, consider
\begin{equation}
\label{phi3}
\Phi_N (x) \propto
\int d\mu {\cal D}[\chi ] \,\delta \left( \int^1_0 d\tau \, \chi _{\tau}
\right)
\exp \left\{ \mu x |g'_N (0)| - \int^1_0 d\tau \, \left[
\frac{1}{2} \dot{\chi }^2_{\tau} + \frac{\mu}{N} | \chi _{\tau} | ^N \right]
\right\} \, .
\end{equation}
The attentive reader may object to the factor $i$ in (\ref{expphi}) being
dropped as well, since $\Phi_N$ should be real. A careful analysis will
reveal that the functional determinant carries this factor, i.e., the
contour
for $\mu$ near the saddle point will be parallel to the imaginary axis.

{}From (\ref{phi3}), we first seek the saddle point, to be denoted by
$\{ -M, X_{\tau} \}$ in $\{ \mu , \chi _{\tau} \}$ space. Note that we
have anticipated this point to be on the negative real $\mu$ axis, so
that $M$ will turn out to be a real and positive. Setting to zero the
variation
of the exponent in (\ref{phi3}) with respect to $\mu$ and
$\chi _{\tau}$ , we have the equations for
the paths which dominate the path-integral:
\begin{equation}
\label{dM}
N x |g'_N (0)|  =  \int^1_0 d\tau \, | X_{\tau} | ^N \, ,
\end{equation}
and
\begin{equation}
\label{dchi}
\ddot{X}_{\tau} + M \, {\rm sgn}(X_{\tau}) \, | X_{\tau} | ^{(N - 1)}
\,
= \, 0 \, .
\end{equation}
Note that, in order to ignore the singularity at $X = 0$, we must restrict
our attention to $N > 1$ here.
The latter being an equation of motion for a classical particle
of unit mass moving in a potential $V(X) = \frac{M}{N}| X | ^N$,
these can be solved easily. The importance of $M>0$ is also now
evident, since periodic $\chi _{\tau}$'s would have been otherwise
impossible. Finally, the constraint
$ \int^1_0 d\tau \, \chi _{\tau} = 0$
can be satisfied trivially.

Integrating (\ref{dchi}) once, we have
\begin{equation}
\label{energy}
\frac{1}{2} \, \dot{X}^2_{\tau} + \frac{M}{N}\, | X_{\tau} |^N = \epsilon
\, ,
\end{equation}
where $\epsilon$ is a constant, representing the total energy in the
mechanical analogy. Assigning $\tau = 0$ to
the point of maximum amplitude (and so, zero velocity), we set
\begin{equation}
\label{bcsonX}
\epsilon  = \frac{M}{N} X^N_0  \, .
\end{equation}
where $X_0 > 0$. Equation (\ref{energy}) can be integrated again in the
usual
manner, exploiting both the periodic boundary conditions and the constraint.
Clearly, we need to focus on only a quarter of the period, so that
\begin{equation}
\label{quarter}
\frac{1}{4} = X_0 ^{(2-N)/2} (N/2M)^{1/2}
\int^1_0 d\xi (1-\xi^N)^{-1/2}\, ,
\end{equation}
where $\xi_{\tau} \equiv X_{\tau}/X_0$. The integral is proportional to
the beta function $B(\frac{1}{N},\frac{1}{2})$, which we will
denote simply as $B$. Now we have a relation between $M$ and $X_0$:
\begin{equation}
\label{XM}
X_0 ^{(N-2)} = \frac {8 B^2} {NM} \, .
\end{equation}
Inserting this into Eqn. (\ref{dM}), eliminating $d\tau$ in favour of
$dX/\dot{X}$ and using (\ref{energy}), we have
\begin{eqnarray}
\label{xM}
N x |g'_N (0)| & = & 4 \int^{X_0}_0 ( dX / \dot{X} ) \, | X | ^N  \\
& = & \left( \frac {8 X_0^{N+2}}{MN}\right)^{1/2}
B(1+ \frac {1}{N}, \frac {1}{2} ) \, .
\end{eqnarray}
Using $B(1+\alpha, \beta) = B(\alpha, \beta) \frac{\alpha}{\alpha + \beta}$,
we find both $X_0$ and $M$ in terms of the scaling variable $x$ :
\begin{equation}
\label{Xx}
X_0 = \left\{ N (\frac {N}{2}+1) x |g'_N (0)| \right \} ^ {1/N}
\end{equation}
and\begin{equation}
\label{Mx}
M = \left ( \frac {8 B^2} {N} \right )
\left\{ N (\frac {N}{2}+1) x |g'_N (0)| \right \} ^ {(2-N)/N} \, .
\end{equation}

With all parameters of the saddle point explicitly determined,
we evaluate the exponential in (\ref{phi3}), i.e., the total ``action".
The result is
$-MN x |g'_N (0)|/2$, so that the leading asymptotic behaviour of $\Phi
_N$
is
\begin{equation}
\label{leadasym}
\Phi_N (x) \sim \exp \left\{ - {\cal S}_N x ^{2/N} \right\} \, ,
\end{equation}
\begin{equation}
\label{classact}
{\cal S}_N = 4 B^2 \left\{ N (\frac {N}{2}+1) \right \} ^ {(2-N)/N}
 |g'_N (0)|^{(2/N)}  \, .
\end{equation}
For $N=2$, $B=\pi$ and, from (\ref{gprime}), $|g'_N (0)|=1/24$, so that
we recover the result in (\ref{phi2}).

Next, we turn to the computation of the prefactor, for which it is
necessary to study the Gaussian fluctuations about the saddle point
$(-M, X_{\tau})$.

To accomplish this, we write $\mu = -M + \hat{\mu}$ and
$\chi _{\tau} = X_{\tau} + \hat{\chi}_{\tau}$ in
(\ref{phi3}) and shift the integration to
$( \hat{\mu} ,\hat{\chi}_{\tau} )$.
Expecting the integrals to be dominated by small
$( \hat{\mu} , \hat{\chi}_{\tau} )$, we expand the argument of
the exponential to second order in these fluctuations. The zeroth order
is
given above in (\ref{leadasym}) while the first order vanishes by the
choice of the saddle point. At the second order, the result is the quadratic
form
\begin{equation}
\label{quadact}
-\hat{\mu} \int^1_0 d\tau \, {\rm sgn}(X_{\tau}) |X_{\tau}|^{(N - 1)}
\hat{\chi}_{\tau} +
\frac{1}{2} \int^1_0 d\tau \, \hat{\chi}_{\tau} {\cal M}
\hat{\chi}_{\tau} \, ,
\end{equation}
where ${\cal M}$ is the operator
\begin{equation}
\label{em}
{\cal M} = -\frac{d^2 \ }{d\tau ^2} - M (N - 1) \, |X_{\tau}|^{(N - 2)}
\, .
\end{equation}
Note that, strictly speaking, we should impose $N \ge 2$, so that the
singularity at $X=0$ can be ignored here also.

Before carrying out the integration over
$( \hat{\mu} ,\hat{\chi}_{\tau} )$,
we discuss several important points.

Firstly, note that ${\cal M}$ is a
hermitian, Schr\"{o}dinger-like operator. It can be diagonalised, and,
since
our problem is based on a finite interval with periodic boundary conditions,
it has a real, discrete spectrum with real, periodic eigenmodes.
Thus, $\hat{\chi}$ can be expanded in terms of these modes and the
the functional integral $\int {\cal D}[\hat{\chi}]$ can be defined as
over the
amplitudes in this expansion.  We will demonstrate that there is a
single zero eigenvalue, associated with a zero-mode, which must be handled
with some care. Defining ${\cal M}'$ to be the operator (\ref{em})
{\it restricted to the subspace orthogonal} to this mode, we
will argue that the spectrum of ${\cal M}'$ is positive,
so that its inverse is well defined and the Gaussian integration in
(\ref{quadact}) over these modes is simple. Furthermore, the product
$M|X_{\tau}|^{(N - 2)}$ in (\ref{em}) can be written
as $(MX_{0}^{(N - 2)})|\xi_{\tau}|^{(N - 2)}$. But, according to
(\ref{XM}), this quantity is independent of $x$, so that the spectrum
of both ${\cal M}$ and ${\cal M}'$ are also $x$-independent. Thus,
the result of the Gaussian integration, which involves ${\rm det}{\cal
M}'$,
is also $x$-independent. Since we are dropping all proportionality
constants in this study, this factor will be neglected.

Secondly, we will show that the zero-mode is also absent from the
off-diagonal $ \hat{\mu}$ - $ \hat{\chi}$ part. As a result, at the
quadratic level, the $ \hat{\chi}$ integration leads to a distribution
for $\hat{\mu}$ of the form exp $( {\cal P} \hat{\mu}^2 /2 )$, where
\begin{equation}
\label{p}
{\cal P} =
\int d\tau d\tau' \, |X_{\tau}|^{(N - 1)} {\cal M}'^{-1} |X_{\tau'}|^{(N
- 1)}
\end{equation}
is positive. To make sense of performing such an integral
(over $\hat{\mu}$), recall that the contour in the $\hat{\mu}$-plane is
parallel to the imaginary axis, by definition of the inverse Laplace
transform. We have simply chosen to have it run
through a saddle point on the real axis: $\mu=-M$. Thus, it is entirely
consistent to choose $\hat{\mu}$ to be pure imaginary, so that
not only is the Gaussian integral well defined, but also the right hand
side of (\ref{phi3}) is now real. Further, we may again use
$X_{\tau}=X_{0} \xi_{\tau}$ to extract the sole dependence of
${\cal P}$ on $x$, given that neither ${\cal M}'$ nor $\xi$ are
functions of $x$. As a result, the integration over $\mu$ leads to
a factor proportional to
\begin{equation}
\label{intmu}
1/\sqrt{{\cal P}} \, \propto \, X_0^{(1-N)} \, .
\end{equation}

Thirdly, for Schr\"{o}dinger problems in one dimension,
the energy levels are non-degenerate, {\it unless} its potential
is a constant. The implication for us is that,
except for the $N = 2$ case, the spectrum of ${\cal M}$
is non-degenerate. Further, since our interval is finite, the spectrum
is discrete, with eigenvalues increasing monotonically with the
number of nodes in the eigenfunction. Now,
the constraint $\delta \left( \int^1_0 d\tau \, \chi _{\tau} \right)$
must clearly be satisfied by both $\hat{\chi}$ and the eigenfunctions.
Therefore, the latter must have an even number ($2n \ge 2$) of nodes.
We will show that the zero-mode has the lowest allowed $n$.
Since it is not degenerate, the rest of the spectrum of ${\cal M}$,
i.e., the spectrum of ${\cal M}'$, must be positive.

Now, let us provide some details of the zero-mode. Its origin is the
translational invariance in our problem. To identify it explicitly,
differentiate (\ref{dchi}) once and find
\begin{equation}
\label{zeromode}
{\cal M} \, \frac{dX_{\tau}}{\ d\tau} = 0 \, .
\end{equation}
Thus, $\dot{X}_{\tau}$ is an eigenfunction of ${\cal M}$ with
zero eigenvalue, so that it is clearly the zero-mode. Next, to show that
it is also absent from the first term in (\ref{quadact}), we note that
${\rm sgn}(X_{\tau}) |X_{\tau}|^{(N - 1)}$ is actually
$dV(X)/dX$. Therefore, if $\hat{\chi}_{\tau}$ is
$\frac{dX_{\tau}}{\ d\tau}$, then the integrand is a perfect
derivative of a periodic function and this term vanishes. The
conclusion is that the zero-mode is completely absent from the
quadratic from (\ref{quadact}). Thus, for the Gaussian approximation
to be valid this mode has to be treated separately. The standard
technique to deal with this situation is the method of collective coordinates
\cite{Raj}. The symmetry here is translational invariance of
(\ref{dM}) and (\ref{dchi}), which implies that the particle
(in the mechanical analogy) can be at the point $X_0$ at any other time
$\tau _0 \in [0,1]$. Thus, there is a one-parameter
family of solutions, labelled by $\tau _0$, all of which satisfy the
differential equation (\ref{dchi}) and the constraint
$ \int_{0}^{1} d\tau \, \chi _{\tau} = 0$.
Indeed, the zero-mode $\dot{X}_{\tau}$ can be recognized as the
difference between $X_{\tau}$ and $X_{\tau - \tau _{0}}$, to lowest
order in $\tau_0$. Using this parameter instead of the amplitude of
$\dot{X}_{\tau}$ in the normal mode expansion of $\hat{\chi}$ enables
us to sum over this component of $\hat{\chi}$, even though it is
entirely absent from the weights.

Having made these remarks, we may now evaluate the Gaussian integral over
$\hat{\chi}$. In principle, we would expand $\hat{\chi}$ in terms
of all the eigenfunctions of ${\cal M}$, except the zero-mode (associated
with the index $n = 0$ below). Therefore,
$\hat{\chi}_{\tau}=\sum_{n\neq 0}a_n \hat{\chi}^{(n)}_{\tau}$.
In the collective coordinates method \cite{Raj},
the integral over the functions $\hat{\chi}$ is replaced by integrals
over $\tau _0$ (which ``replaces" $\a_0$) and $\{a_n | n\neq 0\}$.
The Jacobian of this transformation
is simply $\langle \dot{X} | \dot{X} \rangle ^{1/2}$,
which is proportional to $X_0$ and carries the sole $x$ dependence
here. Now, the integration $\tau_0$ is trivially unity. Meanwhile the
integration over the rest of the
modes ($\{a_n | n\neq 0\}$) has already been
discussed, in connection with the first remark above. No $x$ dependence
appears here. Finally, the integration over $\hat{\mu}$ leads to
(\ref {intmu}). Summarising, the Gaussian integrals yield a prefactor
for
(\ref {leadasym}) proportional to $X_0^{(2-N)}$. Using (\ref {Xx}),
we have our final result:
\begin{equation}
\label{largex}
\ln \Phi_N (x) \sim  - {\cal S}_N x ^{2/N} -
\frac {N-2}{N} \ln(x) + O(1) \ \ \ (x \gg 1) \, ,
\end{equation}
where, explicitly,
\begin{equation}
\label{SN}
{\cal S}_N = \frac{4\pi}{3N(N+2)}
\left(\frac {\Gamma (1/N)}{\Gamma (\frac {N+2}{2N})} \right) ^2
\left\{ \frac {(N+2) \Gamma ( \frac {N+1}{2} )}{2 \sqrt{\pi}} \right\}
^{2/N} \, .
\end{equation}

Although there are some technical limitations to our derivation for
for the $N = 2$ case, they can be circumvented, so that this result
is in fact valid for $N \ge 2$.

\section{The scaling function for small x}

To arrive at $\Phi _N$ in the other limit: $x \ll 1$, we will use
a very different approach, which we will first sketch briefly. As in
the $x \gg 1$ case, we are able to compute only the two leading
terms in $\ln \, \Phi _N$, so that all proportionality constants will
be
dropped. Unlike in the previous section, we will first compute the
asymptotic behaviour of $G_N (\lambda ;T)$ for large $\lambda$
and then rely on a saddle point method to find $\Phi_N$. Our
starting point is the
unscaled version of $G_N (\lambda ;T)$ given by (\ref{genfun}).
So first consider the numerator of the right-hand-side, but without the
constraint $\int^T_0 dt\, \psi _t = 0$, and define
\begin{equation}
\label{alpha}
\alpha \, \equiv \, \lambda /T
\end{equation}
Then,
\begin{equation}
\label{num}
\int {\cal D}[\psi ]\,
\exp - \int^T_0 dt \, \left[
\frac{1}{2} \dot{\psi}^2_t + \alpha |\psi _t| ^N \right]
\, \propto \, \sum_n e^{-E^{(N)}_n T} \, ,
\end{equation}
where $E^{(N)}_n$ is the $(n+1)$-th eigenvalue of the evolution operator
which consists of the quantum-mechanical Hamiltonian for a particle
moving in the potential $\alpha |\psi| ^N$.
As $T\rightarrow \infty$, the ground state dominates, leading to the
Feynman-Kac result \cite{Sch} that the right-hand-side of (\ref{num})
can be replaced by $e^{-E_0^{(N)} T}$, with only an exponentially small
error.
Now, the dependence of $E_0^{(N)}$ on $\alpha$ can be extracted through
dimensional analysis alone, since $\alpha$ has dimensions of inverse time
to
the power $\frac{1}{2}(N + 2)$,
while $E_0^{(N)}$ itself has dimensions of inverse time. Thus, we have
$E_0^{(N)} \propto \alpha ^{2/(N+2)}$ and arrive at the
the leading behaviour of the path integral in
(\ref{num}), for $T \gg \alpha ^{- 2/(N+2)}$.

Having outlined the essential idea, let us now be more systematic. First,
to implement the constraint $\int^T_0 dt \, \psi _t = 0$,
we write the usual integral representation for a delta function.
Next, instead of normalising the path integral as in (\ref{genfun}),
we use $G_2 (\lambda ;T)$ as the denominator. This avoids the complications
from the $\alpha = 0$ system, which lacks a discrete spectrum for a simple
application of the Feynman-Kac result. On the other hand,
$G_2 (\lambda ;T)$ is known explicitly (see (\ref{g2})), so that
the ratio $G_N/G_2$ is essentially our goal.
Thus, we consider, from (\ref{genfun}),
\begin{equation}
\label{ratio_1}
\frac{G_N (\lambda ;T)}{G_2 (\lambda ;T)}  \, = \,
\frac{\int^{\infty}_{-\infty} d\omega \, \int {\cal D}[\psi ]
\, \exp - \int^T_0 dt \, \left[
\frac{1}{2} \dot{\psi}^2_t + \alpha |\psi _t | ^N + i\omega \psi _t
\right] }{\int^{\infty}_{-\infty} d\omega \, \int {\cal D}[\psi ]
\, \exp - \int^T_0 dt \, \left[
\frac{1}{2} \dot{\psi}^2_t + \alpha \psi ^2_t
+ i\omega \psi _t \right] } \, .
\end{equation}
Of course, the presence of $i\omega \psi$ is rather unusual.
However, as we will see, its sole effect is to give a
non-trivial prefactor to $G$ in this limit. We first focus on the
denominator, which consists of simple Gaussian integrals. Without the
constraint, the $i\omega \psi$ term is absent, so that we may use
results from the quantum mechanics of a simple harmonic oscillator
to obtain $G_2 \sim e^{-\sqrt{\alpha /2}\, T}$.
Comparing with (\ref{g2}), we see that this is indeed the leading
behaviour. The next leading term, contained in the prefactor, would
arise from the constraint. In this case, it is easily dealt with, by defining
$\psi '_t = \psi _t + i\omega/2\alpha$ and factorizing the
integrand. The $\psi '$ integral gives us the previous asymptotic form,
while integration over $\omega$ produces the desired prefactor:
$\sqrt{\alpha /T}$. For the $N \ne 2$ case, there is no similar
luxury of factorization, of course. But, by reinterpreting the above
steps, i.e., computing the functional integral with the
$i\omega \psi$ term, as finding an $\omega$-dependent ground state
energy: $E_0^{(2)} (\alpha ,\omega )$, a way forward can be seen.
Thus, we write
\begin{equation}
\label{E2alphom}
\int {\cal D}[\psi ] \, \exp - \int^T_0 dt \, \left[
\frac{1}{2} \dot{\psi}^2_t + \alpha \psi ^2_t
+ i\omega \psi _t \right] \,
\propto \, \sum_n e^{- E^{(2)}_n(\alpha ,\omega ) T} \,
\approx \, e^{- E^{(2)}_0 (\alpha ,\omega ) T} \, ,
\end{equation}
with
\begin{equation}
\label{harmonic}
E^{(2)}_0 (\alpha ,\omega ) = E^{(2)}_0 (\alpha ,0)
+ \omega ^2 /4\alpha \, .
\end{equation}
It is now clear that the integral over $\omega$ gives the prefactor.
Another route to the understanding of (\ref{harmonic}) is to consider
adding a real, linear potential: $h\psi$ first. Clearly, the ground
state energy is well defined, allowing us to arrive at the above result
by analytic continuation to pure imaginary $h$.

Generalizing this approach to the $N \ne 2$ case, we write the
numerator of (\ref{ratio_1}) as
\begin{equation}
\label{GN}
\int^{\infty}_{-\infty} d\omega \, \sum_n
e^{- E^{(N)}_n (\alpha ,\omega ) T} \,
\approx \, \int^{\infty}_{-\infty} d\omega \,
e^{- E^{(N)}_0 (\alpha ,\omega ) T} \, .
\end{equation}
Unlike the $N = 2$ result (\ref{harmonic}), $E_0^{(N)} (\alpha ,\omega
)$ does
not have a dependence on $\omega$ which terminates at second order. It
is
clear, however, by expanding the exponentials in the analogue of
(\ref{E2alphom}) for general $N$, that it has the general form
\begin{equation}
\label{pertexp}
E^{(N)}_0 (\alpha ,\omega ) =
\varepsilon ^{(N)}_{0} (\alpha ) \, +
\omega ^{2} \varepsilon ^{(N)}_{1} (\alpha ) \, +
\omega ^{4} \varepsilon ^{(N)}_{2} (\alpha ) \, + \, \ldots
\end{equation}
So since $\omega$ has dimensions of time to the power $-3/2$,
$\varepsilon ^{(N)}_{m} (\alpha )$ has dimensions of time to the
power $(3m - 1)$. Therefore,
\begin{equation}
\label{epsilon}
\varepsilon ^{(N)}_{m} (\alpha )
= \varepsilon ^{(N)}_{m} (1)\alpha ^{-2(3m-1)/(N+2)} \, .
\end{equation}
To see that only the first two terms in this expansion are relevant,
recall that the integrand depends on $E^{(N)}_0 (\alpha ,\omega )T$
only. Changing the variable of integration from $\omega$ to
$\omega ' \equiv T^{1/2} \alpha ^{- 2/(N+2)} \omega$, so that
the quadratic term in (\ref{pertexp}) is independent of $\alpha$ and
$T$, we see that the $m \ge 2$ terms are of order
$1/[\alpha^{2/(N+2)} T]^{m-1}$. Since this is proportional to
$1/\mu^{2(m-1)/(N+2)}$ and we are interested in the $\mu \gg 1$
limit, we are justified in neglecting all terms beyond $m=1$.
Focusing on the first two terms in (\ref{pertexp}), we write
\begin{equation}
\label{twoterms}
E^{(N)}_0 (\alpha ,\omega )T = {\cal E} \alpha^{2/(N+2)} T
\, + \, {\cal E}' (\omega')^2 \, + \, ...
\end{equation}
Note that ${\cal E} \equiv \varepsilon ^{(N)}_0 (1)$ is just the zero
point
energy of a particle
in the potential $|\zeta |^N$ and is, therefore, positive. Meanwhile,
had $\omega$ been pure imaginary, $E^{(N)}_0$  would certainly have been
lowered, implying that ${\cal E}'$ is also positive. We
note, parenthetically, that it can be estimated in our approach,
since it is just the correction to the ground state energy in
second order perturbation theory. Having ${\cal E}' > 0$,
the integral over $\omega '$ is well defined and provides a simple
constant.

Inserting (\ref{twoterms}) into (\ref{GN}), we have
\begin{equation}
\label{numerator}
\int^{\infty}_{-\infty} d\omega \,
e^{- E^{(N)}_0 (\alpha ,\omega ) \, T} \,
\propto \alpha ^{2/(N+2)} T^{-1/2} \,
\exp  \left[ -{\cal E} \alpha ^{2/(N+2)} T \right] \, \left\{ 1 +
O(\alpha ^{-2/(N + 2)}\, T) \right\} \, .
\end{equation}
Using this, together with a similar expression for the denominator,
\begin{equation}
\label{denominator}
\int^{\infty}_{-\infty} d\omega \,
e^{- E^{(2)}_0 (\alpha , \omega ) T}
\propto  \alpha ^{1/2} T^{-1/2} \,
\exp \left[ - \alpha ^{1/2} T /\sqrt{2} \right]
\end{equation}
we find
\begin{equation}
\label{ratio_2}
\frac{G_N (\lambda ;T)}{G_2 (\lambda ;T)} \propto
\frac{\lambda ^{2/(N+2)}}{\lambda ^{1/2}}
\frac{T^{- 2/(N+2)}}{T^{-1/2}} \,
\frac{\exp \left[-{\cal E} \lambda ^{2/(N+2)} T^{N/(N+2)}\right]}
{\exp \left[-\lambda ^{1/2} T^{1/2} /\sqrt{2} \right] }
\{ 1 + O(\lambda T^{N/2})^{- 2/(N+2)} \}
\end{equation}
where we replaced the $\alpha$'s by the original $\lambda/T$'s.
Finally, using the exact expression (\ref{g2}) for $G_2$, we
arrive at
\begin{equation}
\label{GN2}
G_N (\lambda ;T) \, \propto \,
\lambda ^{2/(N+2)} T^{N/(N+2)}
\exp \left[-{\cal E} \lambda ^{2/(N+2)} T^{N/(N+2)}\right]
\{ 1 + O(\lambda T^{N/2})^{- 2/(N+2)} \} \, .
\end{equation}
Alternatively, we can write it in terms of the scaling form
$g_N (\mu )$ introduced in section 2:
\begin{equation}
\label{gN}
g_N (\mu ) \, \propto \, \mu ^{2/(N+2)}
\exp \left[-{\cal E} (\mu /N )^{2/(N+2)} \right]
\{ 1 + O(\mu ^{- 2/(N+2)} ) \} \, .
\end{equation}
This is the form which $G_N$ or $g_N$ takes in the
$\mu \gg 1$ limit. It should be contrasted with the approach used in
the last section, which was justified in the regime where $\mu \ll 1$.
Thus the results of the last section are complementary to those
derived here.

All that remains to find $\Phi _N (x)$ is to take the inverse
Laplace transform of (\ref{gN}) :
\begin{equation}
\label{muint}
\Phi _N (x) \, \propto \,
\int^{i\infty}_{-i\infty} d\mu \, \mu ^{2/(N+2)}\,
\exp \left[ - {\cal E} (\mu /N )^{2/(N+2)} +
\mu x |g'_N (0)| \right] \, .
\end{equation}
Defining $\nu = x^{(N+2)/N} \mu$ gives, for the exponent in (\ref{muint}),
\begin{equation}
\label{nuexp}
x^{- 2/N} \left\{ - {\cal E} (\nu /N )^{2/(N+2)}
+ \nu |g'_N (0)| \right\}
\end{equation}
so that the integral may be evaluated by steepest descent if $x \ll 1$.
Since ${\cal E} > 0$, there is an extremum at a real positive
value of $\nu$, which is of order unity. Thus, the corresponding
value of $\mu$ is of order $x^{- (N+2)/N} \gg 1$, justifying all the
approximations we made above, based on $\mu \rightarrow \infty$.
Carrying through the integration gives the asymptotic behaviour:
\begin{equation}
\label{smallx}
\ln \Phi_N (x) \sim  - {\cal K}_N x ^{-2/N} -
\frac {N+3}{N} \ln(x) + O(1) \ \ \ (x \ll 1) \, ,
\end{equation}
where ${\cal K}_N$ is a positive constant which depends on some
explicitly known functions of $N$ and on ${\cal E}$, the ground
state energy of a quantum mechanical particle of unit mass in a
potential $V(\zeta)=|\zeta|^N$. Unfortunately, the last item is not
known for general $N$, so that we must be content with just an
implicit expression. Nevertheless, conjecturing that it is monotonic
in $N$, we can place a bound on it: $\pi^2/8$, which is the result
for $N \rightarrow \infty$, where the particle is confined to
$|\zeta| \le 1$ by an infinite square well. Since ${\cal E}$ is
$1/\sqrt{2}$ for $N=2$, it is fairly well bounded numerically
even though its precise value is unknown. Finally, we note that
(\ref{smallx}) reduces to the known result given in
ref. \cite{gof} for the $N = 2$ case.

\section{Conclusions}

To summarise, we have generalised the study of width distributions of
a Gaussian path (random walk) to include moments higher than 2.
Unlike the N=2 case, we are unable to find closed form expressions for
the Laplace transform of these  distributions. Nevertheless, we can
compute their asymptotic behaviour, for both large and small arguments:
eqns. (\ref {largex}) and (\ref {smallx}).
Here, we conclude with a few remarks.

Firstly, note that, though the {\it averages} of the even $N>2$ moments
are trivially related to $\langle w^2 \rangle$, their {\it distributions}
are
not so simple. In particular, we believe that there is no way to obtain
the general $\Phi_N$ or $G_N$ from the N=2 result. They contain
different information about the paths. For example, the following two
paths lead to the same $w^2$ but distinct $w^{(N)}$: (i) $\psi_t =
1$ for $0<t<T/2$, $-1$ for $T/2<t<T$ and $0$ elsewhere,
vs. (ii) $\psi_t =
2$ for $0<t<T/8$, $-2$ for $7T/8<t<T$ and $0$ elsewhere. Thus,
only paths with short excursions to large $\psi$'s, which are presumably
rarer, contribute to the large $x$ tail of $\Phi_N$ with large $N$.
Presumably, this is reflected by eqn. (\ref {leadasym}).

Secondly, we have seen, from Feynman's path-integral formulation
of quantum mechanics, that the Laplace transform of the
distributions for the $N$-th absolute central moments
of random walks are intimately related to the propagation
kernels of a quantum mechanical particle confined to a
potential of the form $V(\zeta)=|\zeta|^N$.  As a result,
the behaviour of $\Phi _{N}(x)$ in the large $x$ limit should be
controlled by the properties of $G_{N}(\lambda)$ for small
$\lambda$. We note that the paths which dominate $G_{N}$ here are
the ones near the ``classical path" (\ref {dchi}). Meanwhile, for the
opposite limit, small $x$ and large $\lambda$, the dominant
contribution to $G_{N}$ comes from the ground state, which is, in a
sense, ``the furthest from classical". It is interesting that these
opposing aspects of the quantum mechanical problem find their
way into the opposite ends of the asymptotics.

Thirdly, let us recall that these distribution functions are
supposedly universal, in the language of renormalisation group.
In other words, they should capture the large-scale properties of
a one-dimensional object stabilised by non-zero tension,
independent of the microscopics of the system. Thus, we may
expect the same behaviour from an
interface between different phases, a polymer in $d=2$, or a
simple model of random walk. The last of these is particularly
easy to study, using Monte Carlo simulations. In this connection,
we should caution the reader on the $N \rightarrow \infty$ limit.
Simulations necessarily deal with systems with finite $L$. To
observe universal properties, we must let $L$ approach infinity.
However, this does not commute with $N \rightarrow \infty$,
so that the results above should be used with some care, if large
values of $N$ were to be used in the investigation.

Finally, it is natural to speculate on further generalizations of
this study. Obvious candidates are interfaces in higher
dimensions \cite{dgt1} and those controlled mainly by curvature
terms \cite{curvature}. Both will involve single component fields.
On the other hand, we could investigate random walks imbedded in
higher dimensions, such as physical polymers in 3-d solutions, where
the periodic boundary condition would correspond to ring polymers.
In this case, we would need multi-component fields and an appropriate
generalisation of the concept of ``width". In the simplest scenario,
$P(w^2)$ will be nothing more than products of the distribution in
\cite{gof}. However, it is clear that $P_N$ will be much more
interesting! While the formulation of any of these problems is
trivial, we anticipate that the analysis itself will be be rather difficult.

\section*{Acknowledgements}

We wish to thank the Isaac Newton Institute for Mathematical Sciences,
Cambridge, UK, where this work was begun while the authors were participating
in the programme on ``Cellular Automata, Aggregation and Growth". We are
grateful to the Departments of Physics at the University of Illinois at
Urbana-Champaign, Virginia Polytechnic Institute and State University
and the University of Manchester for hospitality during the later stages.
This work was supported in part by EPSRC grant GR/H40150 and
US-NSF grant DMR-94-19393.

\end{document}